\documentclass[apjl]{emulateapj}
\usepackage{graphicx}
\usepackage{amssymb}
\usepackage{multirow}
\usepackage{color}
\usepackage{wrapfig}
\usepackage{float}
\usepackage{apjfonts}

\usepackage{lineno}
\usepackage{amsmath}
\usepackage{url}

\newif\ifAMStwofonts
\AMStwofontstrue

\makeatletter

\newcommand{\Rmnum}[1]{\expandafter\@slowromancap\romannumeral #1@}
\makeatother

\shorttitle{EVN Astrometry of PSR J0218+4232}

\shortauthors{Du et al.}

\slugcomment{Published in The Astrophysical Journal Letters (Du et al. 2014, ApJL, 782, L38)}

\begin{document}

\title{Very Long Baseline Interferometry Measured Proper Motion and Parallax of the $\gamma$-ray Millisecond Pulsar PSR~J0218$+$4232}

 \author{
Yuanjie~Du\altaffilmark{1,2},~
Jun~Yang\altaffilmark{3},~
Robert~M.~Campbell\altaffilmark{3},~
Gemma~Janssen\altaffilmark{4,5},~
Ben~Stappers\altaffilmark{4}
~and~
Ding~Chen\altaffilmark{1,2}
}

 \altaffiltext{1}{National Space Science Center, Chinese Academy of
   Sciences, No.1 Nanertiao, Zhongguancun, Haidian district, Beijing
   100190, China; duyj@nssc.ac.cn}

 \altaffiltext{2}{Key Laboratory of Electronics and Information Technology for Space Systems, CAS}

 \altaffiltext{3}{Joint Institute for VLBI in Europe, Postbus~2,
   7990~AA Dwingeloo, The Netherlands}

 \altaffiltext{4}{University of Manchester, Jodrell Bank Observatory,
   Macclesfield, Cheshire, SK11~9DL, UK}

 \altaffiltext{5}{ASTRON, the Netherlands Institute for Radio
   Astronomy, Postbus 2, 7990 AA, Dwingeloo, The Netherlands}

\begin{abstract}
PSR~J0218$+$4232 is a millisecond pulsar (MSP) with a flux density
$\sim$0.9 mJy at 1.4~GHz. It is very bright in the high-energy X-ray
and $\gamma$-ray domains.  We conducted an astrometric program using
the European VLBI Network (EVN) at 1.6~GHz to measure its proper
motion and parallax. A model-independent distance would also help
constrain its $\gamma$-ray luminosity. We achieved a detection of
signal-to-noise ratio S/N~$>37$ for the weak pulsar in all five epochs.
Using an extragalactic radio source lying 20 arcmin away from the
pulsar, we estimate the pulsar's proper motion to be
$\mu_{\alpha}\cos\delta=5.35\pm0.05$~mas\,yr$^{-1}$ and
$\mu_{\delta}=-3.74\pm 0.12$\,mas\,yr$^{-1}$, and a parallax of
$\pi=0.16\pm0.09$\,mas. The very long
baseline interferometry (VLBI) proper motion has significantly
improved upon the estimates from long-term pulsar timing
observations. The VLBI parallax provides the first model-independent
distance constraints: $d=6.3^{+8.0}_{-2.3}$\,kpc,
with a corresponding $3\sigma$ lower-limit of $d=2.3$\,kpc.
This is the first pulsar trigonometric parallax measurement based solely
on EVN observations.
Using the derived distance, we believe that PSR~J0218$+$4232 is the
most energetic $\gamma$-ray MSP known to date.
The luminosity based on even our 3$\sigma$ lower-limit distance is
high enough to pose challenges to the conventional outer gap and
slot gap models.

\end{abstract}


\section{Introduction}


PSR~J0218$+$4232 is a pulsar with a spin period of 2.3~millisecond and
a period derivative of 8.0$\times$10$^{-20}$~s\,s$^{-1}$.
This millisecond pulsar (MSP) was first discovered by
\citet{navarro95} using the Lovell telescope. The radio timing
observations also found that it has a low-mass companion
($M\gtrsim$~0.16~$M_\odot$) with an orbital period of 2 days. Optical
observations using the Keck telescope revealed that the companion is a
helium-core white dwarf with a temperature of
$T_\mathrm{eff}=8060\pm150$~K, and a distance constraint of 2.5--4~kpc
is given by white-dwarf modeling \citep{bassa03}.
Note that the distance uncertainty derived by this method is
  difficult to quantify exactly, because it is dependent on the
white-dwarf's mass and optical luminosity, which are both correlated
with the cooling age that is highly uncertain for observations and
theoretical white-dwarf models \citep{bassa03}.

PSR~J0218$+$4232 is also an energetic pulse emitter in X-rays and
$\gamma$-rays. It was a $\gamma$-ray MSP candidate detected by
Energetic Gamma Ray Experiment Telescope \citep{kuiper00}. Soon after \textit{Fermi Gamma-ray Space Telescope} was
launched, its was confirmed as a $\gamma$-ray MSP \citep{msp-sci}. The
X-ray pulsed emission has also been well detected and monitored by
many X-ray telescopes \citep[][]{webb04}.
%
%
Currently, there are approximately 150 rotation-powered pulsars
detected in the X-ray band, and about 50 of those have millisecond
spin periods \citep{becker09}.  PSR~J0218$+$4232 is one of a few
pulsars with a flux $>10^{-5}$ photon cm$^{-2}$\,s$^{-1}$ in the
2~--~10~keV band \citep[e.g.,][]{kuiper02}. An absolute X-ray timing
accuracy of $\sim$200~$\mu$s was achieved by \emph{Chandra}
\citep{kuiper04} and 40 $\mu$s by \emph{XMM-Newton} \citep{webb04}.
%

Astrometric parameters (e.g. position, proper motion, parallax) can be
determined from pulsar timing observations over a time span of several
years. It is still a challenge to measure the times of arrival of its
pulses with a high precision since PSR~J0218$+$4232 has broad profile
and significant ($\sim$50\%) non-pulsed emission
\citep{navarro95}. The best timing solution published to date was
derived from observations at Effelsberg \citep{laz09}. The proper
motion derived from these observations is
$\mu_{\alpha}\cos\delta=+5.1\pm0.3$~mas\,yr$^{-1}$ and
$\mu_{\delta}=-2.3\pm0.7$\,mas\,yr$^{-1}$,
whose uncertainties are better than the previously published values
in \citet{hobbs05}.
Furthermore, due to the occurrence of sources of timing noise in the
time-of-arrival data (e.g., the pulsar's intrinsic spin-down noise,
noise induced by the stochastic gravitational wave background,
dispersion measure (DM) variations)
as well as a relatively short timespan of timing observations that can
lead to covariances with other parameters of the timing model,
the pulsar-timing method can lead to significant
errors in the estimated astrometric parameters.

The proper motion and trigonometric parallax of a pulsar can also be
independently measured with VLBI observations. The high-precision VLBI astrometry has been applied to
many bright slow pulsars \citep[e.g.,][]{cam96, brisken02, brisken03,
  chatterjee09}, and MSPs, such as PSR~B1937$+$21 \citep{1937},
PSR~J0437$-$4715 \citep{0437}, and PSR~B1257$+$12 with three planets
\citep{yan13}. A pulsar distance measured to 0.4\% accuracy has been
recently achieved in the VLBI astrometry of PSR~J2222$-$0137
\citep{del13}.  Astrometric parameters derived from VLBI observations
can further improve the estimation of parameters from long-term pulsar
timing observations, by providing a prior constraint on astrometric
parameters to which the timing analysis is insensitive, but which may
themselves be highly covariant with other parameters uniquely
approachable via timing.  Combining VLBI- and timing-derived
astrometry can contribute to frame ties between the International
Celestial Reference Frame and the dynamical solar-system frame, which
underlie VLBI and pulsar timing, respectively \citep{madison13}.

\begin{figure*}[!tb]
\centering
\includegraphics[angle=-90,width=0.98\textwidth]{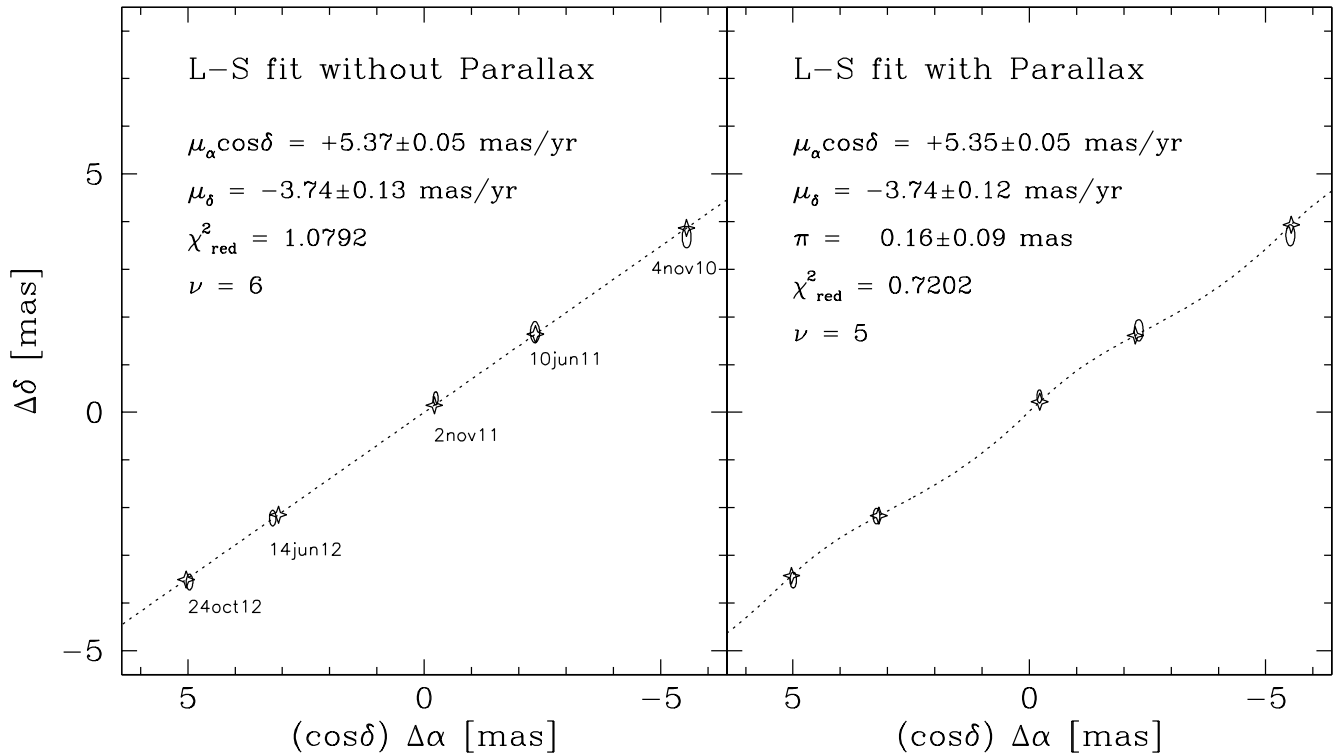}
\caption{EVN astrometry: PSR J0218+4232 positions, $1\sigma$ error
  ellipses, and the modeled track in the sky plane. A star symbol
  denotes the calculated pulsar position at the epoch of each
  observing session. The origin gives the position derived at the
  reference epoch J2011.8793. Some results from the least-squares fit
  are annotated in each panel. Table~\ref{tbl_2} lists the results for
  the fit that includes parallax.  } \label{fig1}
\end{figure*}

In this Letter, we present the results of the first VLBI observations
of PSR~J0218$+$4232. We summarize the strategy of the VLBI
observations and the post-correlation data reduction in Section 2. We
discuss the estimation of the pulsar's astrometric parameters in
Section 3. Finally, we address the model constraints on the
$\gamma$-ray luminosity of the pulsar in Section 4.

\section{VLBI observations and data reduction}

We conducted five epochs of European VLBI Network (EVN) observations of PSR~J0218$+$4232 from 2010 November to 2012 October.  The first three columns of
Table~\ref{tbl_1} describe the basic characteristics of each epoch:
observing date, participating telescopes (abbreviations listed in the
note to the table), and recording bandwidth.  All observations used
2-bit Nyquist sampling in both right- and left-circular polarizations
(thus 1024~Mbps spans 128~MHz on the sky per polarization).  The
central frequency for each epoch was 1658.49~MHz.  Each epoch lasted
6~hr.  The remaining columns in Table~\ref{tbl_1} describe imaging
results, which we will discuss later.

The EVN observations were done in the phase-referencing mode with a
cycle time $t_{\rm cyc}\sim6$~minutes, with a cycle typically comprising
80~s on the phase-reference source and 260~s on the target pulsar,
plus gaps for system-temperature measurements.  We used J0216$+$4240
as the main phase-reference source. This lies 20 arcmin away from the
pulsar and has a compact-double structure with a total flux density
$154\pm15$ mJy at 1.6~GHz.  Secondary phase-reference sources included
J0228+4212 and J0222+4302 (3C66A), which were observed for a single
1-minute scan every 15 minutes.  In our first epoch, we measured the
position of the brighter northern component of the main
phase-reference source
($\alpha_\mathrm{J2000}=+02^\mathrm{h}16^\mathrm{m}25.\!\!^\mathrm{s}972281$,
$\delta_\mathrm{J2000}=+42\degr40^\prime36.\!\!^{\prime\!\prime}42857$)
with respect to J0222$+$4302.
The 1$\sigma$ absolute position uncertainty for J0222$+$4302 is
$\sim0.05$~mas from the GSFC geodetic solution 2009a \citep{NASA09}.
We continued to use this northern component in J0216$+$4240 as the
reference point to track the relative position variation of the pulsar
in all five epochs.  Despite its proximity to the pulsar on sky, it
was still too far to gain the benefits of using it as an in-beam
calibrator.  The FWHM (the full width at half maximum of the
  synthesized beam) of the several 32m telescopes in the EVN would be
about 4\% less than the separation, and the larger, more sensitive
telescopes would certainly need to be slewed. The 76m Jodrell Bank
Lovell telescope (Jb1) has slewing limitations ($\sim$12 source
changes per hour) that require adjustments in the phase-reference
scheduling when the $t_{\rm cyc}$ is under 10 minutes.  The standard
tactic is to omit Jb1 from every other phase-reference source scan,
which in the present case yielded an apparent $t_{\rm cyc}$ of
$\sim$12~minutes for it.

Correlation was performed using the EVN software correlator at JIVE
(SFXC).  The position of the pulsar used for correlation was
$\alpha_\mathrm{J2000}=+02^\mathrm{h}18^\mathrm{m}06.\!\!^\mathrm{s}3580062$,
$\delta_\mathrm{J2000}=+42\degr32^\prime17.\!\!^{\prime\!\prime}379133$.
Pulsar gating during correlation is a technique in which only a
specified portion of the pulse period is actually correlated,
typically providing a signal-to-noise ratio (S/N) gain on the order of the inverse square-root of the pulsar's duty cycle (i.e., assuming no off-pulse
emission).  Gating was not applied during these correlations because
of the pulsar's broad pulse and significant off-pulse emission
\citep{navarro95}.

The data were calibrated using the NRAO Astronomical Image Processing
System \citep{gre03}. The correlation amplitude was
calibrated initially with measured system temperatures and antenna
gain curves for telescopes having reliable data, and with nominal
system equivalent flux density values for other telescopes. The
ionospheric delay was corrected via the total electron content
measurements from GPS monitoring. Phase contributions from the antenna
parallactic angle were removed before fringe fitting. The phase
bandpass solutions were determined from all the calibrator data.
%
%
%
  Some stations had bandpasses characterized by a sharp cut-off
  $\sim1.5$~MHz from the band-edge on each side.  Since the astrometry
  is based primarily on the phases, we did not apply amplitude
  bandpass solutions to avoid introducing noise from these band-edge
  channels when spectral averaging across the bands.
%
The calibrator was imaged in Difmap \citep{she94}, using the
observations in the final epoch, which was the most sensitive one.
The phase contribution arising from the calibrator source structure
was subtracted via re-running the fringe fitting. We did not see any
significant flux or structure variations in the calibrator over the
epochs.
The pulsar position was derived in Difmap \citep{she94} through
fitting the visibility data to a point-source model. Data from Jb1 in
the fourth epoch were ultimately omitted: the combination of its
longer apparent $t_{\rm cyc}$ ($\sim720$\,s) arising from its slewing
limitations as discussed above and higher residual phase rate in this
epoch ($\sim$2.5~mHz) led to a less reliable phase-connection.

Columns 4--10 of Table~\ref{tbl_1} characterize the imaging results of
the five epochs.  Columns 4--6 show the geometry of the
naturally-weighted restoring beam.  Columns 7--8 give the total flux
density and dynamic range (S/N) of the pulsar images.  Columns 9--10 list
the residual position of the pulsar with 1$\sigma$ formal errors
$\frac{\mathrm{FWHM}}{2\,\mathrm{S/N}}$.
We typically achieved an image sensitivity of $\sim$12~$\mu$Jy beam$^{-1}$
(natural weighting) in the epochs having a total bandwidth of 1024
Mbps.  PSR~J0218$+$4232 was detected with a S/N $>37$ in each of the
five epochs.


\begin{table*}[]
\centering
\caption{Summary of the EVN Observations and Imaging Results of the
  PSR J0218$+$4232 \label{tbl_1}}
{ \tiny  
\begin{tabular}{clcccccccc}
\hline \hline
MJD       & Participating Telescopes           & Bandwidth       &  $\theta_\mathrm{maj}$
                                                                           & $\theta_\mathrm{min}$
                                                                                  & $\theta_\mathrm{pa}$
                                                                                           & $S_\mathrm{tot}$
                                                                                                   & $S_\mathrm{peak}/\sigma_\mathrm{rms}$   & $\Delta\alpha\cos\delta$
                                                                                                                           & $\Delta\delta$     \\
(day)      &                                                       & (Mbps)  & (mas)  & (mas)& (deg)   & (mJy)&       & (mas)         &  (mas)            \\
\hline
55504.811  & Ef, Wb, On, Tr, Mc, Sv, Zc, Bd, Jb2                   & 512     &  19.0  & 8.03 & 4.79    & 0.61 & 42.5  & 0.65$\pm$0.09 &  $-$0.26$\pm$0.22 \\
55722.223  & Ef, Wb, On, Tr, Mc, Sv, Zc, Bd, Jb1, Ur, Sh           & 1024    &  16.8  & 7.07 & 7.16    & 0.45 & 37.6  & 3.86$\pm$0.09 &  $-$2.25$\pm$0.22 \\
55867.834  & Ef, Wb, On, Tr, Mc, Sv, Zc, Bd, Jb1, Ur, Sh           & 1024    &  14.0  & 5.33 & 7.61    & 0.59 & 51.2  & 5.96$\pm$0.05 &  $-$3.64$\pm$0.14 \\
56092.209  & Ef, Wb, On, Tr, Mc, Sv, Zc, Bd, Jb1, Ur, Sh, Nt       & 1024    &  14.7  & 6.21 & 7.69    & 0.57 & 42.5  & 9.41$\pm$0.07 &  $-$6.15$\pm$0.16 \\
56224.917  & Ef, Wb, On, Tr, Mc, Sv, Zc, Bd, Jb2, Ur, Sh, Nt, Ro70 & 1024    &  15.8  & 6.58 & $-$15.5 & 0.51 & 47.8  & 11.2$\pm$0.07 &  $-$7.49$\pm$0.16 \\
\hline
\end{tabular}
}
\tablecomments{\scriptsize Ef: Effelsberg, Wb: Westerbork Synthesis Radio Telescope (WSRT) in phase-array mode, On: Onsala 25 m, Tr: Torun, Mc: Medicina, Sv: Svetloe, Zc: Zelenchukskaya, Bd: Badary, Jb1: Jodrell Bank Lovell telescope, Jb2: Jodrell Bank Mark2 telescope, Ur: Urumqi, Sh: Shanghai, Nt: Noto, Ro70: Robledo 70 m}

\end{table*}


\section{EVN Astrometry of PSR~J0218$+$4232}


\begin{table}[h]
\centering
\caption{VLBI astrometry of the PSR J0218$+$4232. \label{tbl_2}}
%
\begin{tabular}{ll}
\hline \hline
Fitting parameter        & Best-fit value \\
\hline
R.A., $\alpha_0$ (J2000)             & $02^{\mathrm h}18^{\mathrm m}06.\!\!^{\mathrm s}358569 \pm 0.\!^{\mathrm s}000002$\\
Decl, $\delta_0$ (J2000)            & $42\arcdeg32^\prime17.\!\!^{\prime\!\prime}37515\pm0.\!^{\prime\!\prime}00008$ \\
Proper motion in R.A., $\mu_\alpha\cos\delta$  & $5.35\pm0.05$\,mas\,yr$^{-1}$ \\
Proper motion in Decl, $\mu_\delta$            & $-3.74\pm0.12$\,mas\,yr$^{-1}$ \\
Parallax, $\pi$                   &  $0.16 \pm 0.09$\,mas   \\
\hline
\end{tabular}
\end{table}

Using least-squares (L-S) minimization, we estimated the standard five
astrometric parameters for the pulsar from its observed residual
positions.  These five parameters are the J2000 position at the
reference epoch 2011.7893 ($\alpha_0$, $\delta_0$), proper motion
($\mu_\alpha\cos\delta$, $\mu_\delta$), and parallax
($\pi$). Table~\ref{tbl_2} lists the results of the L-S fitting.  The
reduced chi-square of this fit was 0.72 (for five degrees of freedom).
The right-hand panel of Figure~\ref{fig1} plots the $1\sigma$ error
ellipses for the pulsar's position at each epoch and its motion on the
plane of the sky computed from the estimated proper motion and
parallax.

PSR~J0218$+$4232 is moving toward the south-east at a
position angle $\sim$125$\arcdeg$. The magnitude of its proper
motion is $\mu_\mathrm{tot}=6.53\pm0.08$~mas\,yr$^{-1}$.  Our VLBI
proper motion estimate is consistent at a 2$\sigma$ level with the
most recently published values from timing observations from
\citet{laz09} ($\mu_\alpha\cos\delta=+5.1\pm0.3$~mas\,yr$^{-1}$ and
$\mu_\delta=-2.3\pm0.7$~mas\,yr$^{-1}$), while the precision of the
EVN estimate is a factor of six better.

Our marginally significant measurement of the parallax,
$\pi=0.16\pm0.09$~mas, corresponds to a distance of
$d=6.3^{+8.0}_{-2.3}$~kpc, and
to a 3$\sigma$ lower-limit to $d$ of 2.3~kpc.  Currently, there is no
distance estimation available from timing observations, making this is
the first model-independent distance measurement.  The uncertainty of
the parallax estimation is reasonable, compared with astrometric
results using a reference source at a similar angular separation
\citep{del13}.  Because the significance of the parallax-parameter
estimate in the L-S fit is only at the $2\sigma$ level, we
also fit a model without parallax.  The left-hand panel of
Figure~\ref{fig1} shows the results of this fit.  It can be seen that
the estimated proper-motion components are insensitive to exclusion of
parallax from the model.  Furthermore, we can integrate the
$F$-distribution formed by the ratio of the chi-square
to variate with the appropriate number of degrees of freedom,
resulting from the two fits in order to characterize the significance
of the parameter ($\pi$) in one of the models \citep[e.g.,][]{bev69,
  kendall94}.  Such a test shows that we can reject the exclusion of
the parallax parameter at only a 34\% confidence level, this is also
consistent with the VLBI data not being able to constrain $\pi$
strongly.  The transverse velocity
derived from the proper motion and parallax is $V_{\rm
  T}=4.74\mu_\mathrm{tot}/\pi=195^{+249}_{-71}$~km\,s$^{-1}$, with the
uncertainty dominated by $\sigma_\pi$.  The statistical mean value of
$\bar{V}_{\rm T}$ for MSPs as a whole is $87\pm13$~km\,s$^{-1}$
\citep{hobbs05}. It seems that PSR~J0218$+$4232 has a large transverse
velocity among MSPs.

%
A model-dependent distance for a pulsar can be derived from its
DM and a Galactic electron density ($n_{\rm e}$) model.
A set of distances derived from VLBI trigonometric parallaxes can
provide point calibrations to models of Galactic $n_{\rm e}$.  In the
case of this MSP, with a DM of 61.25~pc\,cm$^{-3}$ \citep{navarro95},
the TC model \citep{taylor93} yields a distance of 5.7\,kpc and the
NE2001 model \citep{cordes02} a distance of 2.7\,kpc. Thus, the
distance obtained from the TC model is more consistent with our
parallax-based distance in this case.
%


The Doppler correction to the spin-down
luminosity $\dot{E}$ of PSR~J0218$+$4232 is small. The apparent period
derivative $\dot{P}_{\rm shk}$ due to the ``Shklovskii effect''
\citep{shklovskii70} is
\begin{equation}
\dot{P}_{\rm shk} = \frac{V_{\rm T}^2 P}{dc} = 1.5 \times 10^{-21},
\nonumber
\end{equation}
where $c$ is the speed of light, $P$ the rotation period of a
pulsar. $\dot{P}_{\rm shk}$ is thus only a small fraction
($\sim$1.9\%) of the total
$\dot{P}=7.74\times10^{-20}$~s\,s$^{-1}$. In addition, we estimated
the $\dot{P}_{\rm gal}$ induced by Galactic rotation, and it is much
smaller than $\dot{P}_{\rm shk}$. Therefore, both $\dot{P}_{\rm shk}$
and $\dot{P}_{\rm gal}$ can be ignored, and the intrinsic spin-down
$\dot{P}$ dominates the total period derivative.

%


The uncertainties of the estimated pulsar position at the reference
epoch in Table~\ref{tbl_2} are statistical errors from the fit.  The
phase connections between J0216$+$4240 and PSR~J0218$+$4232 worked
quite well and there was no significant red noise (e.g., striping) in
the clean image. This is also as expected from the high declination
and quite small separation (20~arcmin) between these two sources.
Thus, the main systematic position error arises from the phase
connections between J0222$+$4302 and J0216$+$4240 separated by
$1.^\circ2$.  The absolute position error of J0222$+$4302 is
$1\sigma \,= \, 0.05$~mas from the geodetic VLBI solution GSFC2009a
\citep{NASA09}. Using J0216$+$4240 as the reference source, we also
checked the variation of J0222$+$4302 position in these multi-epoch
experiments.
The resulting uncertainty in the derived J0222$+$4302 position, mainly
caused by the ionosphere propagation effect, is $1\sigma$ = $0.2$~mas.
Combining these two factors in quadrature, the measurement of the
systematic position error at the reference epoch is
$1\sigma=0.21$~mas.

\section{The most luminous $\gamma$-ray MSP?}
\begin{figure}[b]
\centering
\includegraphics[angle=0,width=0.48\textwidth]{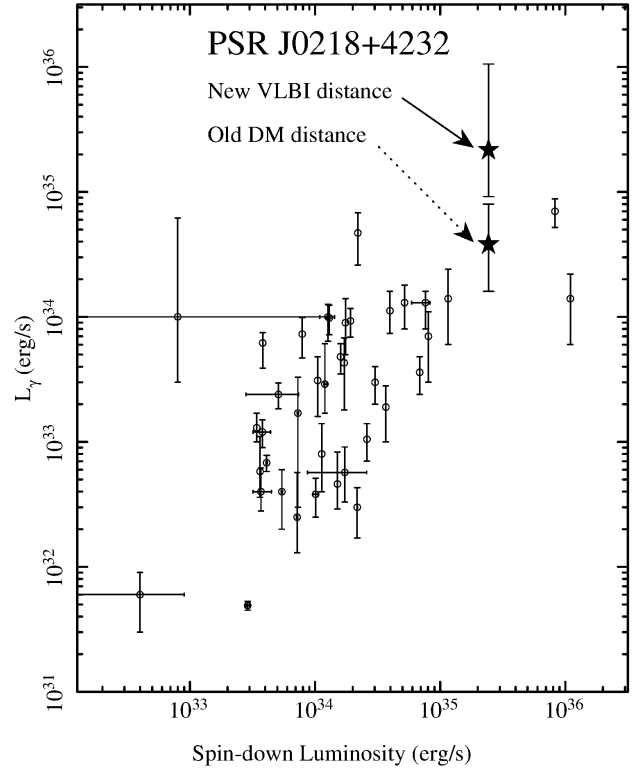}
\caption{$\gamma$-ray luminosity $L_\gamma$ in the 0.1$-$100 GeV
  energy band vs. spin-down luminosity $\dot{E}$. The data, except
  for PSR~J0218$+$4232 (star symbol), are adopted from the second catalog 
  of \emph{Fermi}-LAT MSPs \citep{fermi_cat13} and are
  re-plotted here as a reference sample. Our new 1$\sigma$ parallax
  measurement supports that PSR~J0218$+$4232 is the most energetic
  $\gamma$-ray MSP. }
\label{fig2}
\end{figure}

There are 40 MSPs in the second Fermi Large Area Telescope
Catalog of $\gamma$-ray pulsars \citep{fermi_cat13}. PSR~J0218$+$4232
is one of a few strong $\gamma$-ray MSPs. Its $\gamma$-ray (0.1$-$100
GeV) energy flux is
$F=4.56\times10^{-11}$~erg\,s$^{-1}$\,cm$^{-2}$. \citet{fermi_cat13}
calculated its luminosity, $L_{\rm \gamma}=4\pi f_{\rm \Omega}Fd^2
\approx 3.8 \times 10^{34}$~erg\,s$^{-1}$, based on a DM distance
$d=2.6$~kpc and a beaming fraction $f_{\rm \Omega}=1$.

In Figure~\ref{fig2}, we re-plot $\gamma$-ray luminosity $L_\gamma$
versus spin-down luminosity $\dot{E}$ of the 40 MSPs.  Using our
parallax-based distance (6.3\,kpc) and assuming $f_{\rm \Omega}=1$,
PSR~J0218$+$4232 has a luminosity $L_{\rm \gamma} \approx 2.2 \times
10^{35}$~erg\,s$^{-1}$, a factor of six higher than the previous
measurement. This makes it the most energetic $\gamma$-ray MSP. Its
spin-down luminosity is $\dot{E}=2.4\times10^{35}$~erg\,s$^{-1}$.
Combining its $\dot{E}$ and $L_{\rm \gamma}$, the pulsar has a
$\gamma$-ray efficiency $\eta=L_\gamma/\dot{E}\sim 92$\%, which, while
in the reasonable range, is still quite high.  Even if we use the
  3$\sigma$ lower-limit distance, the corresponding luminosity is
  $L_{\rm \gamma, \,3\sigma } \approx 3.0 \times
  10^{34}$~erg\,s$^{-1}$, and PSR~J0218$+$4232 is still one of the
  three most luminous MSPs in the $\gamma$-ray band.

Below, we discuss the high luminosity that we compute with the
  3$\sigma$ lower-limit distance of PSR~J0218$+$4232 in the context of
  three emission models: the outer gap model, the slot gap
model, and the annular gap model.

The outer gap model is a popular high-energy emission model,
which can explain emission mechanism for MSPs
\citep{cheng13}.  He gives a convenient formula with which to estimate the
$\gamma$-ray luminosity $L_{\rm \gamma,\,OG}$ for MSPs:
\begin{equation}
L_{\rm \gamma, \,OG} \simeq f_{\rm gap}^3 \dot{E}, \\  \nonumber
\end{equation}
where \\
\begin{equation}
f_{\rm gap} = 5.5P^{26/21}B_{\rm 12}^{-4/7} \nonumber
\end{equation}
is the fractional size of the outer gap and $B_{\rm 12}=4.29$
\citep{manchester05} is the surface magnetic field in units of
$10^{12}$~G.  When taking the relevant values of PSR~J0218$+$4232 into
the formulae above, the derived $\gamma$-ray luminosity from the 
  outer gap model is $L_{\rm \gamma,\, OG} = 3.9 \times
  10^{33}$~erg\,s$^{-1}$, which is much smaller than $L_{\rm \gamma,
    \,3\sigma }$.
Even if we consider the effects of both the magnetic inclination
angle ($\alpha$) and the emission geometry for the outer gap model as
described in \citep{zhang04} and \citep{jiang14}, 
 the derived $L_{\rm \gamma,\, OG}$ remains smaller than
$L_{\rm \gamma, \,3\sigma }$. For $\alpha$ in the reasonable
  range of $10^\circ-90^\circ$, $L_{\rm \gamma,\, OG}$ is 1.0$-$12.9
  $\times 10^{33}$~erg\,s$^{-1}$, which is still smaller than $L_{\rm
    \gamma, \,3\sigma }$.  To reach a compatible value with $L_{\rm \gamma, \,3\sigma }$ for PSR~J0218$+$4232, an unphysical value of 30$-$2.3 for
  $\Delta\phi/\Delta\Omega$ would be required \citep{jiang14}. Here $\Delta\phi$ and
  $\Delta\Omega$ are the extension angle at azimuthal direction and
  $\gamma$-ray beaming solid angle, respectively.


The slot gap model, developed from the two-pole caustic model
\citep{tpc03}, is another popular model to explain the high-energy
emission from pulsars \citep{sg03,sg04}.
As described by Harding\footnote{See the report of
  http://www2011.mpe.mpg.de/363-heraeus-seminar/Contributions/2Tuesday/morning/AHarding.pdf},
the $\gamma$-ray luminosity can be given by:
\begin{equation}
\begin{split}
L_{\rm \gamma, \, SG} &\approx 2 \times 10^{34} \epsilon_\gamma \dot{E}_{35}^{3/7}
P_{0.1}^{5/7} \Omega_{\rm SG} \;\; {\rm erg \, s^{-1}} \\ \nonumber
&\simeq 0.2\times 10^{34} \epsilon_\gamma \Omega_{\rm
  SG} \, {\rm \;\; erg\, s^{-1}}, 
  \end{split}
\end{equation}
where $\epsilon_\gamma \lesssim 1$ a conversion efficiency factor
parameterizing how much of the primary particle luminosity converts
into $\gamma$-ray ($>100$\,MeV) emission, $\dot{E}_{35}$ is a pulsar's
spin-down luminosity in units of $10^{35} \, \rm {erg \, s^{-1}}$,
$P_{0.1}$ the spin period in units of 0.1~s, and $\Omega_{\rm SG}$ the
solid angle ($\leq \, 4\pi$) of the slot gap. For PSR~J0218$+$4232,
the derived value of $L_{\rm \gamma, \, SG}$ is smaller than $L_{\rm
  \gamma, \,3\sigma }$ even if $\epsilon_\gamma \sim 1$ and
$\Omega_{\rm SG}\sim 4\pi$ are adopted.

The annular gap model is a
new model under development for treating multi-wavelength emission
from pulsars \citep{qiao04,AG}. \cite{qiao07} gave an analytical
formula of the primary particle luminosity ${L}_{\rm prim, \,
    AG}$ to the annular gap model\footnote{There is a typo in
  Equation 6 of Qiao et al. (2007), where a factor of $c^2$ is omitted
  in the denominator of $\Phi_{\rm Max,\,ann}$.}. Using a conversion
efficiency factor $\epsilon_\gamma$ for $>100$\,MeV band, the
$\gamma$-ray luminosity ($L_{\rm \gamma, \, AG}$) of PSR~J0218$+$4232
in the annular gap model is
\begin{equation}
\begin{split}
L_{\rm \gamma, \, AG} &=\epsilon_\gamma L_{\rm prim, \, AG} =
\frac{16\pi^2 \epsilon_\gamma B^2 R^6}{c^3 P^4}(0.26+0.13 \alpha)^2 \\ \nonumber
&\simeq 1.8\times
10^{35} \epsilon_\gamma R_6^6 \, {\rm \;\; erg\, s^{-1}}, 
\end{split}
\end{equation}
where $R_6$ is the radius of a pulsar in units
 of $10^6$\,cm.  In this model,  $L_{\rm \gamma,\, AG}$ can reasonably
fall within the range of allowable luminosities given our $3\sigma$
lower-limit distance.
Note that the $\gamma$-ray flux may be non-uniformly distributed in the emission beam, it could be varying with line of
sight. $f_{\rm \Omega}$ is model-dependent, with $f_{\rm \Omega}\sim$
0.2 -- 0.8 for MSPs in the annular gap model \citep{AG,MSP_AG13}.


Therefore, the high $\gamma$-ray luminosity derived from our
  $3\sigma$ lower-limit distance for PSR~J0218$+$4232 can be well
  explained by the annular gap model. However, the magnitude of the
  luminosity poses challenges for the outer gap model and the slot gap
  model in their current state.  

\acknowledgments
The authors are grateful to the referee for constructive comments.
We appreciate Li Guo for giving us some useful information. Y.J.D. is
supported by China Postdoctoral Science Foundation (Grant
No. 2012M510047) and the National Natural Science Foundation of China
(Grant No. 11303069 and 11373011). The European VLBI Network is a
joint facility of European, Chinese, South African and other radio
astronomy institutes funded by their national research councils.  This
research has made use of NASA Goddard Space Flight Center's
geodetic VLBI solution 2009a, ATNF Pulsar Catalogue and NASA's
Astrophysics Data System.

\end {document}